\documentstyle[epsf,12pt]{article}
\begin{document}
\begin{titlepage}
\begin{center}
\vspace*{3cm}

\begin{title}
\bold {\large 
    BOSE - EINSTEIN EFFECT \\ 
     
     IN MONTE CARLO GENERATORS\footnote{Talk presented by K. Fia{\l}kowski at the 
XXVIIIth Int. Symp. on Multiparticle Dynamics, Delphi 1998.}}
\end{title}
\vspace{2cm}

\begin{author}
\Large {K. FIA{\L}KOWSKI\footnote{e-mail address:
uffialko@thrisc.if.uj.edu.pl},  R. WIT }
\end{author}\\
\vspace{1cm}
{\sl Institute of Physics, Jagellonian University \\
30-059 Krak{\'o}w, ul.Reymonta 4, Poland}
\vspace{3cm}
\begin{abstract}
{We briefly review various methods of implementing the Bose
- Einstein effect into Monte Carlo generators.  The weight methods are discussed
in more detail; in particular, our method employing a clustering algorithm is
applied for the process $e^+e^-\rightarrow W^+W^-$.  New results for the
multiplicity distributions are presented.  }
\end{abstract}
\end{center}
\vspace{2cm}
TPJU-24/98\\
October, 1998\\
hep-ph/9810492
\end{titlepage}
\section{Introduction} 
\par 
Analyzing Bose - Einstein interference
effects~\cite{hbt} in multiparticle production may provide important information
on the source geometry.  However, more detailed investigations of the space-time
development of this process are difficult.  To compare data
with models we should be able to calculate how the assumptions influence the
interference effects.  Only models using Monte Carlo generators allow us to
produce arbitrary distributions and to investigate the effects of event
selections or kinematical cuts reflecting the real experimental conditions.  On
the other side, however, these models use probabilities and not amplitudes.
Thus incorporating quantum interference effects in them is not an easy task.
\par 
In this talk first we list shortly the existing approaches to the
implementation of Bose - Einstein interference effects into Monte Carlo models.
Then we discuss in more details the weight methods, and in particular a new
method based on a clustering algorithm.  We apply it to the process
$e^+e^-\rightarrow W^+W^-$ (which was recently a subject of a hot debate
~\cite{web}) and present new results concerning the multiplicity in this
process.  We conclude with the outlook for further investigations in this field.
\section{Implementation methods} 
The standard discussion of the BE
effect in multiparticle production~\cite{bgj} starts from the classical space-time 
source 
emitting identical
bosons with known momenta.  Thus the most natural procedure is to treat the
original Monte Carlo generator as the model for the source and to symmetrize the
final state wave function~\cite{sul}.  This may be done in a more proper way
using the formalism of Wigner functions~\cite{zha}.  In any case, however, the
Monte Carlo generator should yield both the momenta of produced particles and
the space-time coordinates of their creation (or last interaction) points.  Even
if we avoid troubles with the uncertainty principle by using the Wigner function
approach, such a generator seems reliable only for heavy ion collisions.  It has
been constructed also for the $e^+ e^-$ collisions~\cite{eg}, but localizing the
hadron creation point in the parton-based Monte Carlo program for lepton and/or
hadron collisions is a rather arbitrary procedure, and it is hard to say what
does one really test comparing such a model with data.  
\par 
It seems to be the
best procedure to take into account the interference effects before generating
events.  Unfortunately, this was done till now only for the JETSET generator for
a single Lund string~\cite{ar,ar2,tnr,hr}, and a generalization for multi-string
processes is not obvious.  No similar modifications were yet proposed for other
generators.  
\par 
The most popular approach, applied since years to
the description of BE effect in various processes, is to shift the final state
momenta of events generated by the PYTHIA/JETSET generators~\cite{sjo,ls}.  The
prescription for a shift is such as to reproduce the experimentallly observed
enhancement in the ratio 
\begin{equation} c_2(Q)=\frac{<n>^2}{<n(n-1)>}
\frac{\int d^3 p_1 d^3 p_2 \rho_2 (p_1,p_2) \delta (Q-\sqrt{-(p_1-p_2)^2})}
{\int d^3 p_1 d^3 p_2 \rho_1 (p_1) \rho_1 (p_2) \delta (Q-\sqrt
{-(p_1-p_2)^2})}, \label{eq:ratio} 
\end{equation} 
\noindent which is a function
of a single invariant variable $Q$. The value of this function is close to one for the 
default JETSET/PYTHIA generator. One parametrizes often this ratio by
\begin{equation} c_2(Q)=1+\lambda exp(-R^2 Q^2), \label{eq:factor}
\end{equation} 
\noindent where $R$ and $\lambda$ are parameters interpreted as
the source radius and "incoherence strength", respectively.  
\par 
After performing the shifts, all the CM 3-momenta of final state particles are
rescaled to restore the original energy.  In more recent versions of the
procedure~\cite{ls2} "local rescaling" is used instead of the global one.  In
any case, each event is modified and the resulting generated sample exhibits now
the "BE enhancement":  the ratio ~(\ref{eq:ratio}) is no longer close to one,
and may be parametrized as in ~(\ref{eq:factor}).  
\par 
There is no theoretical justification for this procedure, so it should be regarded as 
an 
imitation rather than implementation of the BE effect.  Its success or failure in 
describing data
is the only relevant feature.  Unfortunately, whereas the method is very useful
for the description of two-particle inclusive spectra, it fails to reproduce
(with the same fit parameters $R$ and $\lambda$) the three-particle
spectra~\cite{ua1} and the semi-inclusive data~\cite{na22}.  This could be
certainly cured, e.g., by modifying the shifting procedure and fitting the
parameters separately for each semi-inclusive sample of data.  However, the
fitted values of parameters needed in the input factor ~(\ref{eq:factor}) used
to calculate shifts are quite different from the values one would get fitting
the resulting ratio ~(\ref{eq:ratio}) to the same form~\cite{fw1}. This
was shown recently in a much more detailed study~\cite{sm}.  Thus it seems to be very
difficult to learn something reliable on the space-time structure of the source
from the values of fit parameters in this procedure.  
\par 
All this has led to the revival of weight methods, known for quite a long 
time~\cite{pra}, 
but
plagued with many practical problems.  The method is clearly justified within
the formalism of the Wigner functions, which allows to represent (after some
simplifying assumptions) any distribution with the BE effect built in as a
product of the original distribution (without the BE effect) and the weight
factor, depending on the final state momenta~\cite{bk}.  With an extra
assumption of factorization in momentum space, we may write the weight factor
for the final state with $n$ identical bosons as \begin{equation} W(p_1,...p_n)=\sum
\prod_{i=1}^n w_2(p_i,p_{P(i)}), \label{eq:weight} \end{equation} \noindent
where the sum extends over all permutations ${P_n(i)}$ of $n$ elements, and
$w_2(p_i,p_k)$ is a two-particle weight factor reflecting the effective source
size.  A commonly used simple parametrization of this factor for a Lorentz
symmetric source is 
\begin{equation} w_2(p,q)=exp[-(p-q)^2R^2/2], \label{eq:wf}
\end{equation} The only free parameter is now $R$, representing the effective
source size.  In fact, the full weight given to each event should be a product
of factors ~(\ref{eq:weight}) calculated for all kinds of bosons; in practice,
pions of all signs should be taken into account.  Only direct pions and the 
decay products of $\rho, K^*$ and $\Delta$ should be taken into account, since for 
other 
pairs much bigger
$R$ should be used, resulting in negligible contributions.  
\section{Weight methods:  problems and solutions} 
\par 
The main problem of the weight methods is
that weights do change not only the Bose - Einstein ratio ~(\ref{eq:ratio}),
 but also many other distributions.  Thus with the
default values of free parameters (fitted to the data without weights) we
find inevitably some discrepancies with data after introducing weights.  
\par 
We
want to make clear that this cannot be taken as a flaw of the weight method.
There is no measurable world ``without the BE effect'', and it makes not much
sense to ask, if this effect changes e.g.  the multiplicity distributions.  If
any model is compared to the data without taking the BE effect into account, the
fitted values of its free parameters are simply not correct.  They should be
refitted with weights, and then the weights recalculated in an iterative
procedure.  This, however, may be a rather tedious task.  
\par 
Therefore we use
a simple rescaling method proposed by Jadach and Zalewski ~\cite{jz}.  Instead
of refitting the free parameters of the MC generator, we rescale the BE weights
(calculated according to the procedure outlined above) with a simple factor
$cV^n$, where $n$ is the global multiplicity of "direct" pions, and $c$ and $V$
are fit parameters.  Their values are fitted to minimize 
\begin{equation}
\chi^2 = \sum_n[cV^nN^w(n) - N^0(n)]^2/N^0(n) 
\label{eq:chi2} 
\end{equation}
where $N^0(n)$ is the number of events for the multiplicity $n$ without weights,
and $N^w(n)$ is the weighted number of events.  This rescaling restores the
original multiplicity distribution ~\cite{fww}.  In addition, the single
longitudinal and transverse momentum spectra are also restored by this rescaling
~\cite{fww}.  
\par 
Obviously, for a more detailed analysis of the final states,
single rescaling may be not enough.  E.g., since different parameters govern the
average number of jets and the average multiplicity of a single jet, both should
be rescaled separately to avoid discrepancy with data.  Let us stress once again
that such problems arise due to the use of generators with improperly fitted
free parameters, and do not suggest any flaw of the weight method.  Another
problem is that our formula for weights~(\ref{eq:weight}) is derived using some
approximations, which are rather difficult to control~\cite{bk}.  We can justify
them only {\it a posteriori} from the phenomenological successes of the weight
method.  
\par 
An obvious simplification introduced above is the Gaussian form of
the two particle weight factor (\ref{eq:wf}).  The detailed form of this factor
should reflect the space time chracteristics of the source.  Thus one may be
forced to introduce asymmetry between various space-time components of momenta
and/or more complicated functions (e.g., separate terms for direct particles and
decay products of various resonances).  
\par 
Last but not least, the main practical difficulty with formula (\ref{eq:weight}) is 
the 
factorial increase of the number of terms in the sum with increasing multiplicity of 
identical pions $n$.  For high energies, when $n$ often exceeds 20, a straightforward
application of formula (\ref{eq:weight}) is impractical~\cite{hay}, and some
authors ~\cite{jz,kkm} replaced it with simpler expresssions, motivated by some
models.  It is, however, rather difficult to estimate their reliability.  
\par
We have recently proposed two ways of dealing with this problem.  One method
consists of a truncation of the sum (3) up to terms, for which the permutation
$P(i)$ moves no more than 5 particles from their places ~\cite{fw2}. However, it is 
difficult to claim a priori that such a truncation does not change the results which 
would 
be obtained using the full series (2).
\par 
Therefore a second way of an approximate calculation of the
sum (2) was proposed ~\cite{jw}.  Since this sum, called a permanent of a matrix
built from weight factors $w_{i,k}$, is quite familiar in field theory, one may
use a known integral representation and approximate the integral by the saddle
point method. However, this method is reliable only if in each row (and column)
of the matrix there is at least one non-diagonal element significantly different
from zero.  Thus the
prescription should not be applied to the full events, but to the clusters, in
which each momentum is not far from at least one other momentum.  The full
weight is then a product of weights calculated for clusters, in which the full
event is divided.
\par 
The considerations presented above suggested the necessity of combining these 
two methods.  After dividing the 
final state momenta of identical particles into clusters, we used for small clusters 
exact 
formulae presented in~\cite{fw1,fw2}.  For large clusters (with more than five 
particles) 
we compared two approximations (truncated series and the integral representation) to
estimate their reliability and the sensitivity of the final results to the
method.  Obviously, the results depend also on the clustering algorithm:  if we
restrict each cluster to particles very close in momentum space, the neglected
contributions to the sum ~(\ref{eq:weight}) from permutations exchanging pions
from different clusters may be non-negligible, and if the cluster definition is
very loose, the saddle point approximation may be unreliable.  This was then
also checked to optimize the algorithm used.  We found that the truncated series
method was sufficient in all cases~\cite{fww}.\\

\section{Example:  $e^+e^- \to W^+W^-$ } 
As already noted, the main problem of
the weight methods is the lack of a reliable reference model for "the world
without the Bose - Einstein effect".  Therefore it is particularly interesting
to use the method for two processes, which are influenced by this effect in
different ways.  An important example is the comparison of two channels of the
process named in the title:  the decay of both $W-s$ into 2-j hadronic states,
and the channel where one $W$ decays into leptons and the other one decays into
hadrons.  
\par 
With the advent of LEPII data one started to discuss in detail
possible effects which could influence the final state obtained from double
hadronic decay and break the simple factorization picture.  The original
motivation was the concern about the possibility of using these data for the
precision measurements of $W$ mass, crucial for the tests of the standard model.
Although the original suggestions of possible large $W$ mass shifts due to the
BE interference and the colour reconnection (CR) effects~\cite{eg} seem to be
rather exaggerated (see refs.~\cite{web,mor}, and references quoted therein),
there are other possible observables which may discriminate between the existing
models of space - time development of the hadronization process.  
\par 
The
simplest observable of this type is just the average multiplicity.  Its value
for the double hadronic $WW$ decay $\overline n_{WW}$ may be not just twice the
average multiplicity from single $W$ hadronic decay $\overline n_W$.  Moreover,
one may predict which one is bigger.  
\par 
For the $WW$ production at the
energies near the threshold the decay products of both $W$-s are formed in
the same space - time region.  Therefore the transition amplitudes to the
hadronic final states are not just products of two decay amplitudes.  The
symmetrization enhances the probabilities to obtain final states where the
momenta of identical pions are close in momentum space.  This is more likely for
higher multiplicities, where many pions are slow in the CM frame.  Thus one may
expect that $\Delta \overline n \equiv \overline n_{WW} - 2\overline n_W > 0$.
This is to be contrasted with the CR effects, which reduce the
multiplicity~\cite{ks}. These two effects may well cancel to large extent (the so
called BE conspiracy~\cite{ls2}).  Thus it is particularly interesting to form
quantitative predictions for each effect, which is quite simple for the BE
effect implemented by our weight method~\cite{fww}.  
\par 
For the MC without the
BE effect the final state of the hadronic $WW$ decay is a simple superposition
of two $W$ decay product systems.  This means that the generating function of
the multiplicity distribution \begin{equation} \label{s7} G(z) = \sum P(n)z^n.
\end{equation} should be just the square of the generating function for a single
decay \begin{equation} \label{s4} G_{WW}(z) = [G_W(z)]^2.  \end{equation} 
\par
Rescaling the distribution $P(n)$ by $cV^n$ factors rescales the argument of $G$
by $V$ (the normalization factor $c$ is irrelevant since any $G(z)$ has to
fulfill the equation $G(1) = 1$).  This does not spoil the relation (4).
Rescaling may be interpreted as refitting the parameter which controls the
density of particles from a single string.  Thus the same value of the rescaling
parameter V should be used for single- and double $W$ decay.  
\par 
In the single
decay this value is fitted to restore exactly the average multiplicity obtained
without weights (and compatible with the data).  However, for the double decay
such a reduction is insufficient:  the BE weight for a final state from the
double decay is always bigger than the product of weights for two independent
decays.  Since weights enhance the probability of states with high multiplicity,
an excess of multiplicity appears.  This should be contrasted with the momentum 
shifting method, in which by definition the multiplicity distributions are unchanged
 by the BE effect.
\par 
To be more precise, the rescaling should be
performed separately for each channel of the process under consideration in
order to avoid changes, e.g., of the $W$ branching ratios into $u\overline d$
and $c \overline s$ channels.  However, we have checked that for our purposes a
single rescaling is sufficient, since the difference of multiplicities obtained
with the two rescaling procedures is much smaller than the estimated uncertainty
of the final result.  
\par 
We performed a quantitative analysis~\cite{jww} with
the value of only free parameter $R$ from the formula (\ref{eq:wf}) determined
by the fit to the Bose-Einstein ratio in the $Z^0$ decay~\cite{hoo}. We found the
excess of the multiplicity $\Delta \overline n = 2.1 \pm 0.9$ in a good
agreement with the average preliminary data from all the LEPII experiments at
$172 ~GeV$~\cite{dap}, which give $\Delta \overline n = 2.4 \pm 1.8$.  These
results seemed to be confirmed by the first data from $183 ~GeV$, where much bigger
statistics is collected~\cite{aleph}.  However, the average of all LEPII
experiments at this energy shows no significant excess of multiplicity:  the
value of $\Delta \overline n$ is $0.2\pm 0.5$~\cite{facet}.  The values from four
experiments fluctuate rather widely around this number, as shown in the table
below.  
\vspace{0.5cm}

~~~~~~~~~~~~~~~\begin{tabular}{||l|c|c|c||} 
\hline & $\overline n_{WW}$ & $\overline n_W$ &
$\Delta \overline n$ \\ \hline \hline ALEPH & 35.3$\pm$0.4$\pm$0.6 &
17.0$\pm$0.3$\pm$0.2 &+1.3$\pm$0.8 \\ \hline DELPHI & 37.4$\pm$0.5$\pm$0.9 &
19.5$\pm$0.4$\pm$0.6 &-1.6$\pm$1.5 \\ \hline L3 & 36.3$\pm$0.4$\pm$0.8
&18.6$\pm$0.3$\pm$0.4&-1.0$\pm$0.9 \\ \hline OPAL & 39.4$\pm$0.5$\pm$0.7 &
19.3$\pm$0.3$\pm$0.3 & +0.7$\pm$1.0 \\ 
\hline 
\end{tabular} 

\vspace{0.5cm}
 
\par
The estimates of the (negative) shift of the multiplicity due to the colour
reconnection (CR) effect are only around $-0.5$~\cite{ks}, and cannot explain
the disagreement of our prediction with data.  There is, however, a good reason
why our prediction could overestimate seriously the multiplicity excess.  
\par
As already noted, we have used a single value of the parameter $R$ (reflecting
the average distance between pions) for all pairs of "direct" pions.  For the
pairs coming from different $W-s$ this is not quite correct already at the
threshold of $WW$ production, since the strings formed in two $W\rightarrow
q\overline q$ decays are not aligned, and only slow pions may be expected to
originate really from the same volume.  Moreover, using a single value of $R$ is
an approximation which becomes worse with increasing energy.  
\par 
Both BE and
CR effects mentioned above are expected to decrease at higher energy.  For the
BE effect a simple approximate method to estimate this decrease would be to
define an effective average distance for pairs of pions from different $W$-s as
given by 
\begin{equation} 
\label{s6} R^{2}_{eff} = R^{2}_{W} + 4\beta^2 \gamma^2
c^2 \tau^2 
\end{equation} 
where $\beta$ and $\gamma$ are the velocity and the
Lorentz factor of each $W$ in the CM frame at given energy.  $R_W$ defines the
effective size of $W$, as measured by the BE effect in the $W$ (or $Z$) decay,
and should be used for pairs coming from the same $W$. The proper time $\tau $ should 
represent the combined effect of the $W$'s lifetime and  the hadronization time for 
slow 
$W$ decay products, which dominate the BE effect. We take for $\tau$ the value of 
0.9fm/c, which gives for the second term in (\ref{s6}) 
the value of $1fm^2$ at $183 ~GeV$ and $0.5fm^2$ at $172 ~GeV$.
\par
To apply the corrected algorithm we should be able to discriminate between pairs of 
pions coming 
from a single $W$ and from two $W-s$. Therefore we have modified our program. In our 
previous 
programs we were calling the procedures from 
the "inside" of JETSET (at the same place, where the original LUBOEI procedure was 
called). Now 
we take the final state produced by JETSET, identify the ancestors of pions and select 
only 
"direct" pions labelling them by the sign of $W$, from which they originate.
\par
We found a strong reduction of the excess multiplicity due to the extra term 
in~(\ref{s6}). 
Whereas without this term we reproduce the previous value of $\Delta \overline n$, for 
the values 
quoted above we obtain $\Delta \overline n = 1.1$ at $172 ~GeV$, and only $0.7$ at 
$183 
~GeV$. Since we neglect the (negative) CR effect, the net excess of multiplicity may 
well be 
negligible for $183 ~GeV$ and higher energies, as the data seem to indicate.
\par
It is rather difficult to estimate the uncertainty of our predictions. Even without 
the second 
term in (8) we found that changes of $R_W$ at the level of 10\% (experimentally 
allowed) lead to 
about 40\% change of $\Delta \overline n$ (\ref{s6}). The value of $\tau$ is even less 
known. 
Allowing for the uncertainty of about 20\%  around the standard value of 1fm/c we find 
a 40\% error 
of the second term in (8) and the corresponding uncertainty of  $\Delta 
\overline n$ between 60 and 80\%. In Fig.1 we show the resulting error band of our 
predictions for 
$\Delta \overline n$ as a function of energy together with the two points representing 
the data 
averages at $172~GeV$~\cite{dap} and $183~GeV$ ~\cite{facet}. Obviously, neither our 
predictions nor 
the data are accurate enough to draw any strong conclusions, but there is a hope of 
confirming or 
disproving the effect in future investigations.\\

\vspace{0.3cm}
\epsfxsize=10cm
~~~~~~~~~~~\epsfbox{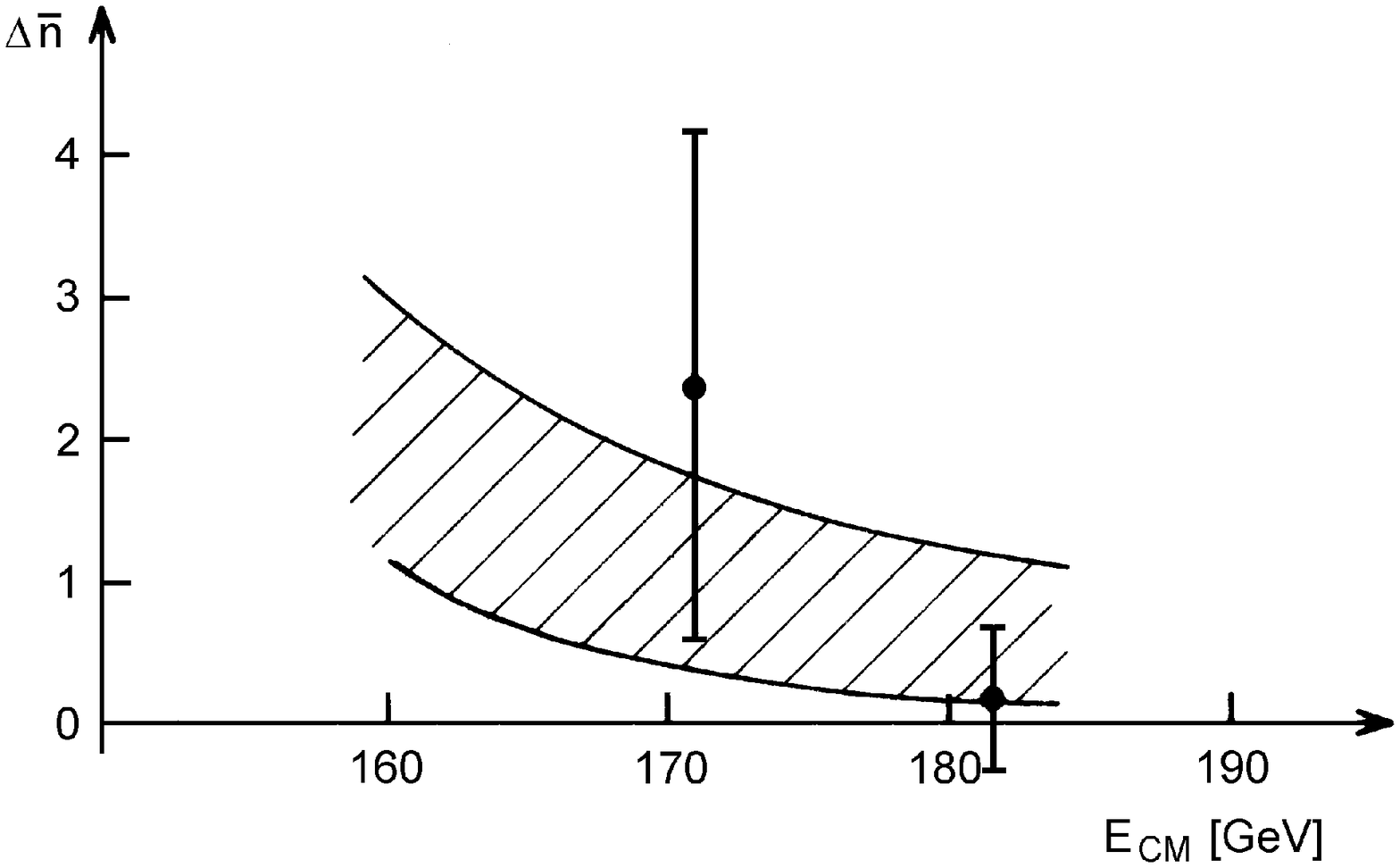}
\vspace{0.2cm}
\par
{\bf Fig.1.}
{\sl The excess of multiplicity $\Delta \overline n$  as a function of CM energy 
$E_{CM}$. 
Shaded area represents our predictions from the BE effect, and the black points with 
errors 
are the averages of the LEP-II data at two energies.}
\vspace{0.5cm}
\par 
A more direct way to investigate the BE effect for the joint $WW$ decay is
to subtract the distributions for the joint $WW$ decay and for the single $W$
decay to get the separated BE effect for pairs from two different $W$-s.
However, this has not given yet conclusive results~\cite{hoo,facet}.  Some preliminary
data suggest that at $172 ~GeV$ the value of $R_{eff}$ for such pairs should be
similar to that of $R_W$, whereas other data show a supression instead of BE
enhancement at low Q (!).  Certainly better data are needed to clarify the
situation.  In any case the conclusions drawn from the multiplicity
excess/deficit and from the low $Q$ enhancement/suppresion should be compatible.
\section{Conclusions and outlook} 
We have seen that our weight method of
implementing the Bose - Einstein effect into Monte Carlo generators seems to
work quite well.  Moreover, for the process $e^+e^-\to W^+W^-$ it gives
distinctly different predictions than the momentum shifting
method~\cite{sjo,ls,ls2}.  Thus in principle it is possible to decide on 
the basis of data
which method is better.  
\par 
There are obvious directions to extend the
applications of our method.  The range of very small $Q^2$, for which the
"intermittency signals" were observed, should be analysed in more detail,
possibly using non-gaussian weight factors (instead of (\ref{eq:wf})).  The
possibility of non-symmetric sources, represented by non-symmetric weight
factors should be considered.  Semi-inclusive data should be investigated,
looking for the possible dependence of the size parameter $R$ on different
variables.  Finally, higher order effects should be analysed too.

\section*{Acknowledgments} The financial support from KBN grants No 2 P03B 086
14 and No 2 P03B 010 15 is gratefully acknowledged.

\end{document}